\documentclass[usenatbib]{mn2e}
\usepackage{natbib}
\usepackage{epsf}
\usepackage{graphicx}
\def\oabs{[O~\small III\large ] }
\def\ofig{[O~\small III] }

\title[The jet-powered optical nebula of Cygnus X--1]{The jet-powered optical nebula of Cygnus X--1}
\author[D. M. Russell et al.]
{D. M. Russell$^1$\thanks{Email: davidr@phys.soton.ac.uk},
R. P. Fender$^1$, E. Gallo$^{2,3}$ and C. R. Kaiser$^1$
\\
$^1$School of Physics \& Astronomy, University of Southampton, Highfield, Southampton, SO17 1BJ, UK\\
$^2$Department of Physics, University of California, Santa Barbara, CA 93106, USA\\
$^3$Chandra Fellow\\
}

\begin{document}
\maketitle

\begin{abstract}
We present H$\alpha$ and \oabs (5007 \AA) images of the nebula powered by the jet of the black hole candidate and microquasar Cygnus X--1, observed with the 2.5m Isaac Newton Telescope (INT). The ring-like structure is luminous in \oabs and there exists a thin outer shell with a high \oabs / H$\alpha$ flux ratio.  This outer shell probably originates in the collisionally excited atoms close to the front of the bow shock. Its presence indicates that the gas is shock excited as opposed to photoionised, supporting the jet-powered scenario. The shock velocity was previously constrained at 20 $< v_{\rm s} < 360$ km s$^{-1}$; here we show that $v_{\rm s} \geq 100$ km s$^{-1}$ (1$\sigma$ confidence) based on a comparison of the observed \oabs / H$\alpha$ ratio in the bow shock with a number of radiative shock models. From this we further constrain the time-averaged power of the jet: $P_{\rm Jet} = (4$--$14)\times 10^{36}$ erg s$^{-1}$. The H$\alpha$ flux behind the shock front is typically $4\times 10^{-15}$ erg s$^{-1}$ cm$^{-2}$ arcsec$^{-2}$, and we estimate an upper limit of $\sim 8\times 10^{-15}$ erg s$^{-1}$cm$^{-2}$arcsec$^{-2}$ (3$\sigma$) to the optical ($R$-band) continuum flux of the nebula. The inferred age of the structure is similar to the time Cyg X--1 has been close to a bright H~\small II\large~region (due to the proper motion of the binary), indicating a dense local medium is required to form the shock wave. In addition, we search a $> 1$ degree$^2$ field of view to the south of Cyg X--1 in H$\alpha$ (provided by the INT Photometric H$\alpha$ Survey of the Northern Galactic Plane; IPHAS) for evidence of the counter jet interacting with the surrounding medium. Two candidate regions are identified, whose possible association with the jet could be confirmed with follow-up observations in [S~\small II\large ] and deeper observations in \oabs and radio.

\end{abstract}


\section{Introduction}

Galactic black holes release an unknown fraction of their accreting matter and energy in the form of collimated outflows (or jets) that travel into the surrounding medium \citep[for a review see][]{fend06}. These accreting black hole candidate X-ray binaries (BHXBs) spend a large fraction of their lifetimes in a `hard' X-ray state \citep{homabe05,mcclet06} which is ubiquitously associated with the formation of partially self-absorbed synchrotron emitting radio jets \citep*[e.g.][]{fendet04}. While the presence of these jets is commonly inferred from a flat radio-through-infrared spectral component, radio observations of the nearby \citep*[d = 2.1$\pm$0.1 kpc;][]{masset95} high-mass BHXB Cygnus X--1 have directly resolved a collimated jet $\sim$30 AU in length \citep{stiret01}, whilst in this X-ray state. A transient radio jet has also been observed of length $\sim$140 AU that was launched during a period of X-ray state transitions \citep{fendet06}. Transient jets with high Lorentz factors are sometimes seen when a BHXB leaves the hard state \citep[e.g.][]{fendet04}.

\begin{figure}
\centering
\includegraphics[width=7.37cm,angle=0]{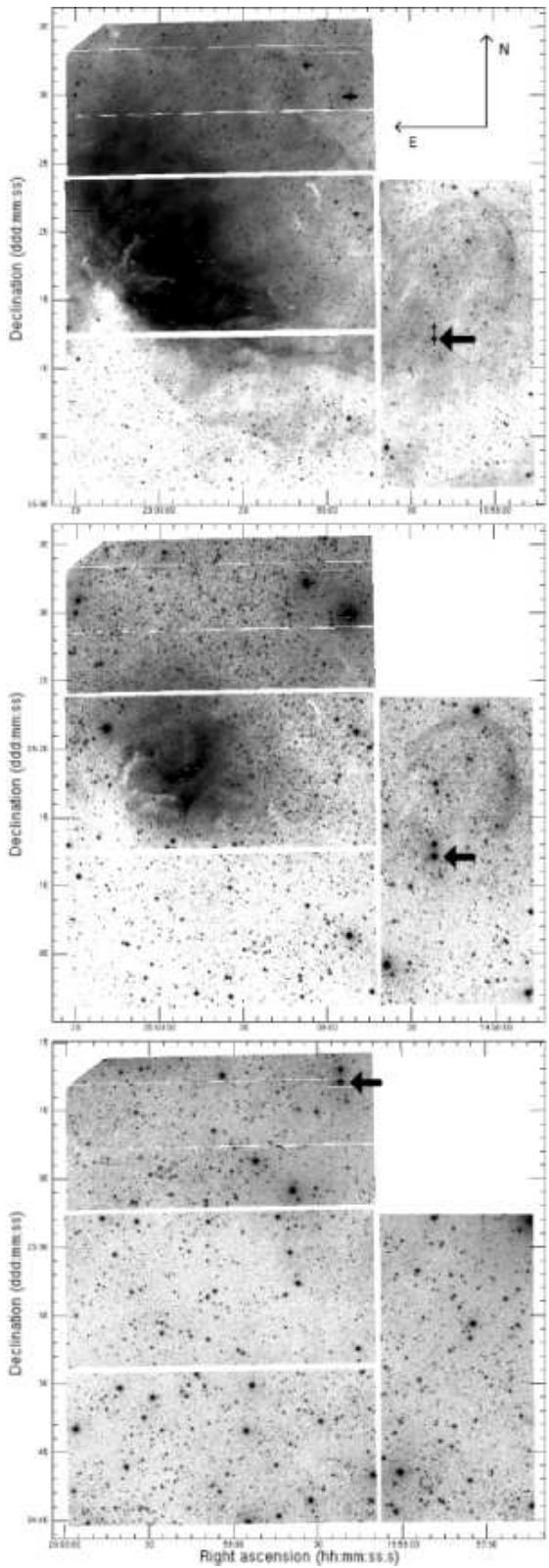}
\caption{Stacked WFC images of the field of Cyg X--1. Top: 2400 sec in H$\alpha$ (6568\AA); Centre: 2400 sec in \ofig (5007\AA); Lower: 1800 sec \ofig (5007\AA) of the field south of Cyg X--1. Cyg X--1 is indicated by an arrow in each panel. In the top and middle panels, the H~\small II region Sh2--101 is prominent on the eastern CCD chips and the Cyg X--1 shell nebula is visible to the north west of the system.
}
\end{figure}

The power carried by the self-absorbed, steady hard state jets when averaged over their lifetime stands as one of the key questions in our understanding of BHXB accretion, and possibly for black hole accretion on all mass scales. In recent years it has become apparent that a significant fraction of the liberated accretion power is ejected as outflows in hard state BHXBs \citep*[e.g.][]{gallet03}, and it now seems likely that in hard X-ray states the accretion flow is `jet-dominated' \citep*{fendet03}, with more power in the jets than in X-rays. Clearly, measuring as accurately as possible the jet power is key to understanding both the overall physics of the accretion process and the matter and energy input from BHXBs into the interstellar medium (ISM).

Attempts at estimating the jet power content from core radio luminosities are riddled with assumptions about its spectrum and radiative efficiency, the latter of which is poorly constrained \citep[e.g.][]{ogleet00,fend01,homaet05,hein06}. The jet power estimated in this way is highly sensitive to the location of the high-frequency break of the flat (spectral index $\alpha \approx 0$ where $F_{\nu}\propto \nu^{\alpha}$) optically thick part of the jet spectrum, as the radiative power is dominated by the higher energy photons \citep{blanko79}. This break is likely to be at $\sim 2 \mu$m, at least for BHXBs at high luminosities ($L_{\rm X}\ga 10^{36}$ erg s$^{-1}$) in the hard state \citep{corbfe02,nowaet05,russet06}.

The jet power may alternatively be constrained by analysing its interaction with the surrounding medium, without requiring prior knowledge of the jet spectrum and radiative efficiency. Radio lobes associated with jets from Active Galactic Nuclei (AGN) are commonly used as accurate calorimeters of the \emph{power $\times$ lifetime} product of the jets \citep{burb59}, a method only very recently applied to jets from stellar mass black holes. Radio lobes have been identified and associated with an increasing number of BHXBs \citep{miraet92,rodret92,corbet02} and even a couple of neutron star X-ray binaries \citep*{fomaet01,tudoet06}.

Recently, deep radio observations of the field of Cyg X--1 resulted in the discovery of a shell-like structure which is aligned with the aforementioned resolved radio jet \citep{gallet05}. This \emph{jet-blown nebula} with a diameter of $\sim 5$ pc is interpreted as the result of a strong shock that develops at the location where the collimated jet impacts on the ambient ISM. The nebula was subsequently observed at optical wavelengths by the Isaac Newton Telescope (INT) Wide Field Camera (WFC) and is clearly visible in H$\alpha$ (6568 \AA) line emission, and coincident with the radio shell. The H$\alpha$ flux density $F_\nu \geq 0.02$ mJy arcsec$^{-2}$ is $\geq 20$ times the measured radio flux density, inconsistent with optically thin synchrotron emission between the two spectral regimes and instead favouring a thermal plasma \citep{gallet05}.

\begin{table*}
\begin{center}
\small
\caption{List of observations with the INT WFC.}
\begin{tabular}{lllllll}
\hline
Run no.&MJD&Target&RA \& Dec. (centre of &Airmass&Filter&Exposure\\
       &   &      &CCD 4; J2000)         &       &      &time (sec)\\
\hline
473434&53653.922&Cyg X--1 north&19 59 38.49 +35 18 20.0&1.171&\ofig 5007 \AA&1200\\
473435&53653.937&Cyg X--1 north&19 59 39.31 +35 18 29.6&1.231&\ofig 5007 \AA&1200\\
473436&53653.956&Cyg X--1 north&19 59 38.50 +35 18 20.0&1.310&Harris V  &600 \\
473437&53653.963&Cyg X--1 north&19 59 38.49 +35 18 20.0&1.380&H$\alpha$ 6568 \AA&1200\\
473438&53653.978&Cyg X--1 north&19 59 39.31 +35 18 30.0&1.494&H$\alpha$ 6568 \AA&1200\\
473439&53653.996&Cyg X--1 north&19 59 37.68 +35 18 10.0&1.620&Harris R  &600 \\
473441&53654.013&Cyg X--1 south&19 59 07.99 +34 56 59.9&1.860&\ofig 5007 \AA&900 \\
473442&53654.025&Cyg X--1 south&19 59 08.80 +34 57 09.8&2.064&\ofig 5007 \AA&900 \\
473452&53654.106&Landolt 94--171&02 53 38.80 +00 17 17.9&1.139&Harris R  &10\\
473454&53654.109&Landolt 94--171&02 53 38.79 +00 17 18.0&1.138&Harris V  &10\\
473457&53654.113&Landolt 94--702&02 58 13.39 +01 10 54.0&1.128&Harris V  &10\\
473458&53654.113&Landolt 94--702&02 58 13.39 +01 10 53.9&1.128&Harris R  &10\\
\hline
\end{tabular}
\normalsize
\end{center}
The positions refer to the centre of CCD 4, which is the middle of the three eastern CCDs (Fig. 1). 10 bias frames and 5$\times$H$\alpha$, 7$\times$\ofig 5007 \AA, 6$\times$Harris $R$-band and 5$\times$Harris $V$-band sky flats were taken in the same observing night.
\end{table*}

\begin{figure*}
\centering
\includegraphics[width=17.7cm,angle=0]{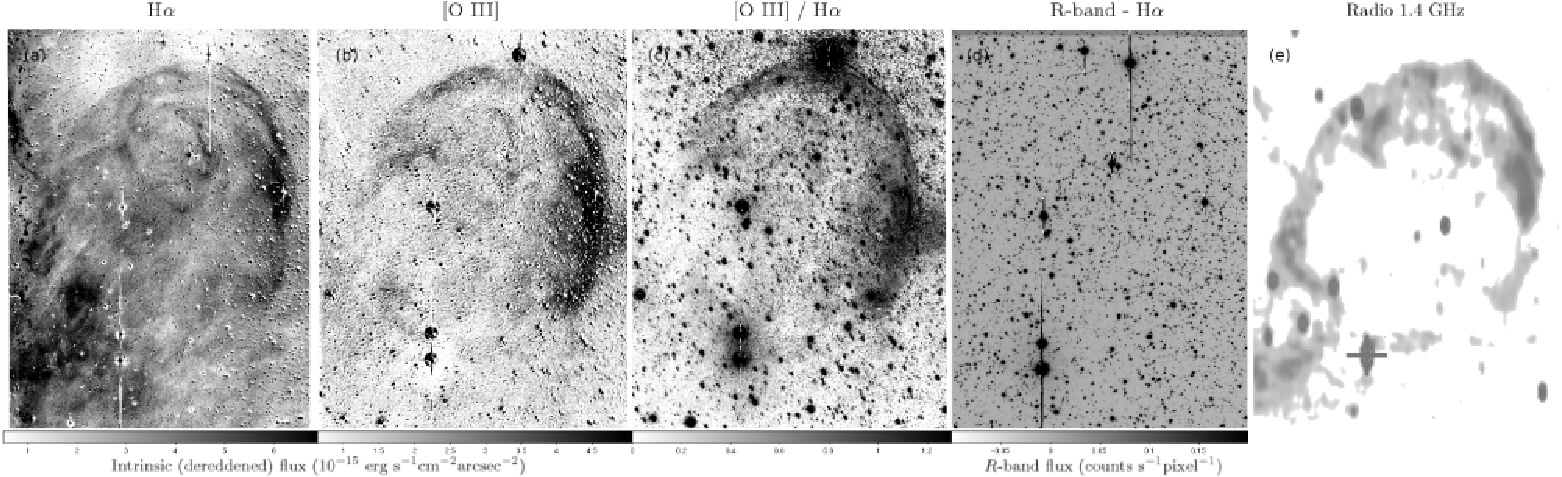}
\caption{Images of the Cyg X--1 ring nebula in: (a) H$\alpha$; (b) \ofig; (c) \ofig / H$\alpha$; (d) $R$-band - H$\alpha$ and (e) radio 1.4 GHz \citep{gallet05}. The H$\alpha$ and \ofig images are de-reddened using $A_{\rm V}$ = 2.95 measured for Cyg X--1 \citep{wuet82}, flux calibrated and continuum-subtracted (i.e. H$\alpha-R$ and \ofig$-V$). The \ofig / H$\alpha$ flux ratio image is created from the non-continuum-subtracted \ofig and H$\alpha$ frames. The H$\alpha$ contribution has been subtracted from the $R$-band image on the right, in which the nebula is not visible. Cyg X--1 is the lower of the two bright stars near the bottom of each image and is marked by a cross in the radio image.
}
\end{figure*}

Models of X-ray binary jet--ISM interactions predict a shell of shocked compressed ISM emitting bremsstrahlung radiation containing a bubble of relativistic synchrotron--emitting plasma \citep{kaiset04}. As with any interstellar shock wave where the preshock gas is at least partially ionised by the approaching radiation field, line emission from recombination of the shocked gas is also expected \citep[e.g.][]{cox72}. The shell, which is essentially a radiative shock wave, consists of a bow shock front where the gas is in transition to a higher temperature due to collisions and plasma instabilities \citep[e.g.][]{mckeho80}. The post-shocked gas then enters the optically thin `cooling region', where the overall ionisation level rises then falls as the gas radiatively cools to $\sim 10^4$ K \citep[e.g.][]{coxra85}. Shock waves are commonly observed in supernova remnants (SNRs), and shocks created from outflows exist in Herbig-Haro objects, where emission lines are produced in the shock wave created by bipolar flows from pre-main-sequence stars impacting the ISM \citep[e.g.][]{schw83}. Shock fronts associated with AGN jets interacting with the intra-cluster gas (which is much hotter than the ISM) are also seen at X-ray energies \citep*{cariet94,krafet03,formet05,nulset05,wilset06,croset06}. In addition, infrared sources found symmetric about the position of the X-ray binary GRS 1915+105 \citep{chatet01} may be jet--ISM impact sites \citep{kaiset04}.

By modelling the shell emission of the Cyg X--1 nebula as originating from radiatively shocked gas \citep*{castet75,kaisal97,heinet98}, the total power carried by the jet was estimated to be $\sim 9 \times 10^{35}\leq P_{\rm Jet} \leq 10^{37}$ erg s$^{-1}$ which, when taking into account the power of the counter jet, is equal to $\sim $0.06--1 times the bolometric X-ray luminosity of Cyg X--1 \citep{gallet05}. These calculations have led to estimates of the matter content of the jet \citep{hein06}; a similar technique to those applied to the jets of AGN, where their matter content are inferred from a combination of information from the core and lobes \citep*[e.g.][]{celofa93,dunnet06}.

These jet power calculations are highly sensitive to the velocity of the gas in the shock front. From temperature constraints and from the non-detection of an X-ray counterpart, this was estimated to be $20 \leq v_{\rm s} \leq 360$ km s$^{-1}$. Emission line ratios of shock-heated gas can constrain its parameters, including the velocity of the shock \citep[e.g.][]{oste89}.

Here, we performed wide field imaging observations of the region of Cyg X--1 in H$\alpha$ and [O~\small III\normalsize ] (5007 \AA) filters in order to (a) constrain the velocity and temperature of the shocked gas and hence the time-averaged power of the X-ray binary jet of Cyg X--1, (b) obtain for the first time flux-calibrated optical emission line measurements from a nebula powered by an X-ray binary jet, and (c) search for any jet--ISM interactions associated with the counter jet supposedly to the south of Cyg X--1.

\section{Observations and data reduction}

The field of Cyg X--1 and of two Landolt standard stars were observed using the Wide Field Camera (WFC) on the 2.5m Isaac Newton Telescope (INT) in narrowband and broadband filters. The images were taken on 2005-10-10 as part of a Galactic survey of emission line nebulae powered by X-ray binary jets (Russell et al., in preparation). The conditions were cloudless with a seeing of $\sim$1--2 arcsec. The moon was set, making it possible to achieve a higher signal-to-noise ratio (S/N) than previous images presented in \cite{gallet05}. Table 1 lists the observations used for this work. The WFC consists of 4 CCDs of 2048$\times$4100 imaging pixels, each of scale 0.333 arcsec pixel$^{-1}$. The field of view is 34$\times$34 arcmin.

Data reduction was performed using the pipeline package \small THELI \normalsize \citep[details in][]{erbeet05}. After manually separating the data into type (bias/flat/science) and filter, \small THELI \normalsize de-biased and flat-fielded the science frames. The package read the exposure time and RA \& Dec from the \small FITS \normalsize file headers (it is currently configured for reducing data from $\sim30$ instruments including the INT WFC) and matched several hundred stars in each exposure with those in the online MAST Guide Star Catalog (release 2.2). \small THELI \normalsize then aligned and stacked (for each position and filter) the images, and position-calibrated mosaics were created, with flux in counts s$^{-1}$pixel$^{-1}$. The resulting H$\alpha$ and [O~\small III\normalsize ] mosaics of the Cyg X--1 field are displayed in Fig. 1.

\section{Flux calibration}

The Landolt standard stars 94--171 ($V=12.659$; $R=12.179$) and 94--702 ($V=11.594$; $R=10.838$) were observed at low airmass (see Table 1). We performed aperture photometry of the standards in the $V$ and $R$-band reduced images with \small PHOTOM \normalsize and accounted for airmass--dependent atmospheric extinction according to \cite{king85}. The resulting conversion between intrinsic flux density (at airmass $=0$) and counts s$^{-1}$ differed between the two standards by a factor 0.5 and 6.5 percent for $V$ and $R$, respectively. The H$\alpha$ filter is located within the passband of the Harris $R$ filter, and [O~\small III\normalsize ] is located within the Harris $V$ filter. For a flat spectrum source, the WFC $R$-band filter collects 15.742 times more flux than the H$\alpha$ filter, and likewise $V$ collects 11.127 times more flux than [O~\small III\normalsize ]. By multiplying by these factors we obtain $7.140\times 10^{-3}$ mJy (intrinsic) per counts s$^{-1}$ for H$\alpha$ and $5.637\times 10^{-3}$ mJy (intrinsic) per counts s$^{-1}$ for [O~\small III\normalsize ].

As a check of our calibration, we measured the fluxes of 10 stars in our reduced Cyg X--1 H$\alpha$ image, that are listed in the INT Photometric H$\alpha$ Survey of the Northern Galactic Plane \citep[IPHAS;][]{drewet05} catalogue (which uses the same telescope, instrument and filter). The ratio of the H$\alpha$ flux of each star measured here to that listed in the catalogue is 1.06$\pm$0.39 ($1\sigma$). We can therefore assume our flux calibration is accurate, and we adopt a $1\sigma$ error for each flux measurement $F$ (H$\alpha$ and [O~\small III\normalsize ]), of 0.14 dex (i.e. $log~F \pm 0.14$).

\begin{figure}
\centering
\includegraphics[height=8.5cm,angle=270]{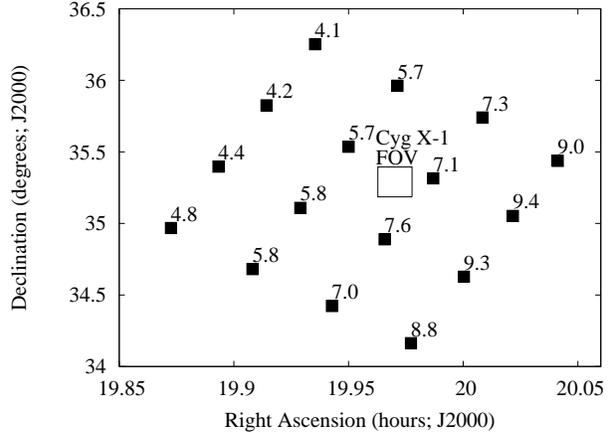}
\caption{Total Galactic neutral hydrogen column density $N_{\rm H}$ in units of $10^{21}$ cm$^{-2}$ at positions close to Cyg X--1 \citep{hartbu97}. The box indicates the field of view (FOV) containing Cyg X--1 and its nebula.
}
\end{figure}

We created flux-calibrated continuum-subtracted H$\alpha$ and [O~\small III\normalsize ] images of the region of Cyg X--1 (Fig. 2a, 2b) using \small IMUTIL \normalsize and \small IMMATCH \normalsize in \small IRAF \normalsize. The resulting images are corrected for interstellar extinction towards Cyg X--1 adopting $A_{\rm V}=2.95\pm0.21$ \citep{wuet82}. The extinction towards Cyg X--1 and its nebula may vary over the field of view. The total neutral hydrogen column density $N_{\rm H}$ through the Galaxy at various positions in a $\sim4$ deg$^2$ area around Cyg X--1 \citep{hartbu97} is indicated in Fig. 3. $N_{\rm H}$ appears to vary smoothly over the field, and the four closest points to Cyg X--1 yield $N_{\rm H}=(6.55\pm 0.94)\times 10^{21}$ cm$^{-2}$ in the region containing Cyg X--1, or $A_{\rm V}=3.66\pm 0.53$ adopting $N_{\rm H} = 1.79 \times 10^{21} cm^{-2}A_{\rm V}$ \citep{predet95}. If the extinction $A_{\rm V}$ towards Cyg X--1 (i.e. not through the whole Galaxy in this direction) also varies by $\pm$0.53 over the much smaller field of view of the nebula (a conservative case), an error is introduced of 0.18 dex and 0.25 dex to the H$\alpha$ and [O~\small III\normalsize ] flux measurements, respectively (no distance estimate is required). The measurements of [O~\small III\normalsize ] / H$\alpha$ however only suffer a 0.06 dex uncertainty due to the different extinctions suffered at the two wavelengths ([O~\small III\normalsize ] at 5007 \AA ~and H$\alpha$ at 6568\AA; this concerns only the contribution to the error budget of the extinction).

Propagating the above two sources of error, we arrive at:
\begin{eqnarray}
  \Delta (log_{10} F_{\rm H\alpha})=0.23~log_{10} F_{\rm H\alpha}\\
  \Delta (log_{10} F_{\rm [O~\small III\normalsize ]})=0.29~log_{10} F_{\rm [O~\small III\normalsize ]}\\
  \Delta (log_{10} (F_{\rm [O~\small III\normalsize ]}/F_{\rm H\alpha}))=0.15~log_{10} (F_{\rm [O~\small III\normalsize ]}/F_{\rm H\alpha})
  \end{eqnarray}
\vspace{0mm}

[O~\small III\normalsize ] / H$\alpha$ and $R$-band $-$ H$\alpha$ images were also created (Fig. 2c, 2d). The [O~\small III\normalsize ] / H$\alpha$ image represents the ratio of the [O~\small III\normalsize ] and H$\alpha$ fluxes (in de-reddened erg s$^{-1}$cm$^{-2}$arcsec$^{-2}$) and the H$\alpha$ contribution is subtracted from the continuum in the $R$-band $-$ H$\alpha$ image. The 1.4 GHz radio image of \cite{gallet05} in shown in Fig. 2e for comparison.

\section{Results}

The H$\alpha$ and [O~\small III\normalsize ] fluxes measured in a number of positions in the shell of the nebula are plotted in Fig. 4. It seems that the [O~\small III\normalsize ] / H$\alpha$ ratio increases with position going clockwise around the shell, from values of $F_{\rm[O~\small III\normalsize ]}$ / $F_{\rm H\alpha}\sim 0.2$ on the eastern side to $\sim 1.2$ on the western side of the nebula. The fluxes of both emission lines are also higher on the western side. The line ratio is approximately constant in the filaments inside the nebula, with values of $F_{\rm[O~\small III\normalsize ]}$ / $F_{\rm H\alpha}\sim 0.7$. The H~\small II \normalsize region Sh2--101 and the `stream' of diffuse emission joining the region to the Cyg X--1 nebula, possess ratios of order $F_{\rm[O~\small III\normalsize ]}$ / $F_{\rm H\alpha}\sim 0$--0.2, with one area (RA 19 59 45, Dec +35 18 48) having a higher value, $F_{\rm[O~\small III\normalsize ]}$ / $F_{\rm H\alpha}\sim 0.6$. This area resembles a shell-like feature close to a bright star HD 227018 at the centre of the H~\small II \normalsize region, and could be a bow shock associated with the motion of this star \citep[which is 12 milli-arcsec per year in the direction of the observed shell in the H~\small II \normalsize region;][]{perret97}. We measure the flux in the H~\small II \normalsize region to be $F_{\rm H\alpha}\sim 5$--$15\times 10^{-14}$ erg s$^{-1}$cm$^{-2}$arcsec$^{-2}$; $F_{\rm[O~\small III\normalsize ]}\sim 0.5$--$5\times 10^{-14}$ erg s$^{-1}$cm$^{-2}$arcsec$^{-2}$.

\begin{figure}
\centering
\includegraphics[width=8.48cm,angle=0]{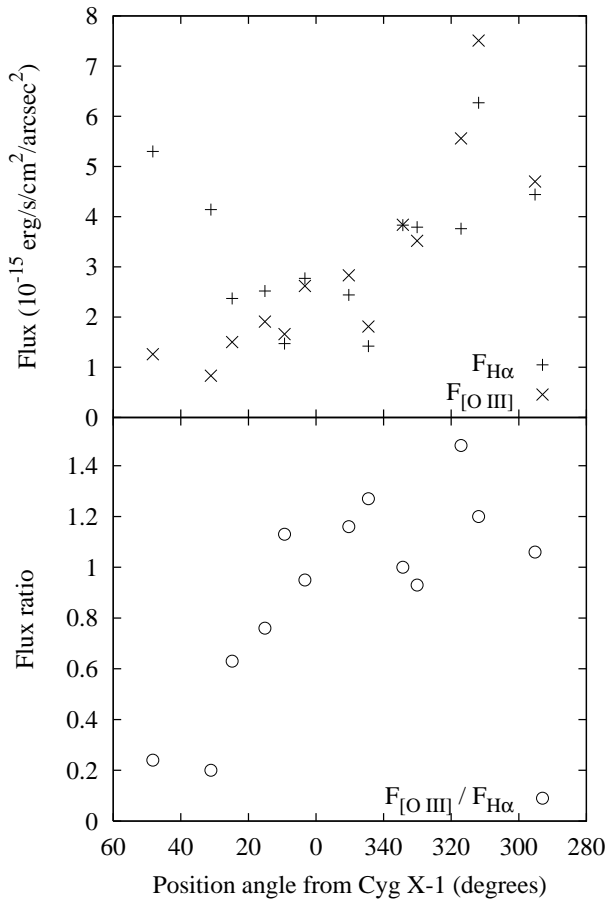}
\caption{Intrinsic de-reddened continuum-subtracted H$\alpha$ and \ofig fluxes at various positions around the shell, versus the position angle from Cyg X--1. See equations 1--3 for the errors associated with each flux measurement. At each position, the mean flux in a 164 arcsec$^2$ circular aperture containing no visible stars is plotted. The data on the left are from the eastern side of the nebula and the data on the right are from the western side.
}
\end{figure}

The large quantity of stars in the line of sight of the shell prevent us from accurately calculating the total H$\alpha$ and [O~\small III\normalsize ] luminosity of the structure. However, we can conservatively estimate the area of the emitting gas to be between 3 arcmin$^2$ (approximating the nebula to a $15\times 700$ arcsec$^2$ area) and 46 arcmin$^2$ (the area of a circle with radius 4 arcmin) and take the mean flux of the nebula (Fig. 4) to constrain its luminosity. At a distance of 2.1 kpc, the H$\alpha$ and [O~\small III\normalsize ] luminosities of the nebula are $1.8\times 10^{34}\leq L_{\rm H\alpha}\leq 2.8\times 10^{35}$ erg s$^{-1}$ and $1.3\times 10^{34}\leq L_{\rm[O~\small III\normalsize ]}\leq 2.1\times 10^{35}$ erg s$^{-1}$, respectively.

In ten regions (of $\sim 5$ arcsec$^2$ areas) inside and outside the H$\alpha$-emitting nebula, the $R$-band de-reddened H$\alpha$-subtracted flux is consistent with zero, and the mean of the standard deviations in each region is $2.57\times 10^{-15}$ erg s$^{-1}$cm$^{-2}$arcsec$^{-2}$. We therefore have a $3\sigma$ upper limit to the $R$-band flux of $7.7\times 10^{-15}$ erg s$^{-1}$cm$^{-2}$arcsec$^{-2}$, or a flux density of 0.0068 mJy arcsec$^{-2}$. Similarly, we obtain a $3\sigma$ upper limit to the $V$-band flux (from ten $\sim 5$ arcsec$^{-2}$ regions) of $2.0\times 10^{-14}$ erg s$^{-1}$cm$^{-2}$arcsec$^{-2}$, or 0.0196 mJy arcsec$^{-2}$. From the $R$-band upper limit, we conclude that the optical continuum flux density is $\leq 7$ times the radio flux density \citep{gallet05}, yielding a radio--optical spectral index $\alpha \leq 0.15$.

The broadband spectrum is consistent with bremsstrahlung emission with a flat radio--optical continuum and optical emission lines which are at least one order of magnitude more luminous than the continuum. We cannot however definitely state that the radio emission is thermal in origin, as mixed thermal/non-thermal plasma is seen in some environments \citep[e.g. the arcsec-scale jets of SS 433;][]{miglet02}. A future measurement of the radio spectral index, which is difficult to achieve, is required to test whether the radio through optical spectrum is consistent with a thermal plasma or requires two (or more) components. Currently the morphology of the radio emission, being spatially coincident with the rather thin optical shell, does seem to favour the single thermal plasma model.

\subsection{Morphology and nature of the nebula}

The higher S/N in the H$\alpha$ image compared to that in \cite{gallet05} is striking from the observed structure within the nebula (Fig. 2a). One would expect the most luminous areas of the nebula to occur in regions of the shock front where the line of sight is tangential to its surface and therefore should exist only at the apparent edge of the nebula if the structure is spherical. Filaments of gas are seen within the nebula and are unlikely to originate in unassociated line-of-sight photoionised gas because of their filamental morphology; the nebulosity is less diffuse than the gas in the photoionised H~\small II \normalsize region. The [O~\small III\normalsize ] / H$\alpha$ ratios of $\sim 0.7$ in the filaments (see above) and morphology suggest either multiple shock fronts or an uneven non-spherical nebula. The latter is expected for a varying preshock gas density, which is clearly the case on inspection of the variation in H$\alpha$ nebulosity close to the nebula.

The [O~\small III\normalsize ] / H$\alpha$ image reveals two striking features: firstly the H$\alpha$ nebulosity (which extends from the H~\small II \normalsize region Sh2--101 to the east of the nebula) possesses a low [O~\small III\normalsize ] / H$\alpha$ flux ratio compared to the Cyg X--1 nebula, and secondly there exists a thin outer shell to the Cyg X--1 nebula with a higher [O~\small III\normalsize ] / H$\alpha$ ratio. H~\small II \normalsize regions and photoionised gas in general can possess a wide range of [O~\small III\normalsize ] / H$\alpha$ ratios \citep*[e.g.][]{john53,blaiet82} but this morphology of an outer [O~\small III\normalsize ]-emitting shell is only expected from shock-excited gas. These two observations confirm that the gas in the Cyg X--1 nebula must be ionised by a different source to the surrounding nebulosity. 

In Fig. 5 we show a blow-up of the north-west area of the nebula in [O~\small III\normalsize ] / H$\alpha$; clearly showing this thin shell of enhanced [O~\small III\normalsize ] / H$\alpha$ ratio on the outer side of the ring. We have taken a 3.3~arcsec-wide slice through the image (indicated in Fig. 5) and plot the emission line flux and $F_{\rm[O~\small III\normalsize ]}$ / $F_{\rm H\alpha}$ flux ratio along this slice in Fig. 6. The upper panel of Fig. 6 shows that there is a sharp outer edge to the ring, and this edge is further out in [O~\small III\normalsize ] than in H$\alpha$. This results in the thin outer shell of enhanced [O~\small III\normalsize ] / H$\alpha$ ratio that is visible in the lower panel of Fig. 6. This outer shell is $\sim 6$ arcsec thick and the rest of the ring is $\sim 40$ arcsec thick, at least in the region of the slice. We note that the thin shell of enhanced [O~\small III\normalsize ] / H$\alpha$ ratio is not due to inaccurate alignment of the [O~\small III\normalsize ] and H$\alpha$ images as the stars are aligned and the thin shell exists on both sides of the nebula.

\begin{figure}
\centering
\includegraphics[width=7.5cm,angle=0]{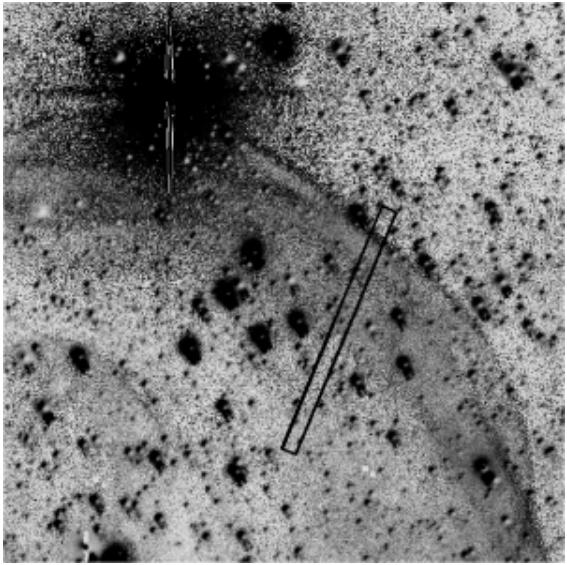}
\caption{A blow-up of part of Fig. 2c: \ofig / H$\alpha$ image of the Cyg X--1 ring nebula. We interpret the thin outer shell with a high \ofig / H$\alpha$ flux ratio as possibly originating in the ionised atoms close to the front of the bow shock. North is up, east to the left. The emission line fluxes and \ofig / H$\alpha$ ratio along the slice indicated in the image are shown in Fig. 6.
}
\end{figure}

\begin{figure}
\centering
\includegraphics[width=8cm,angle=0]{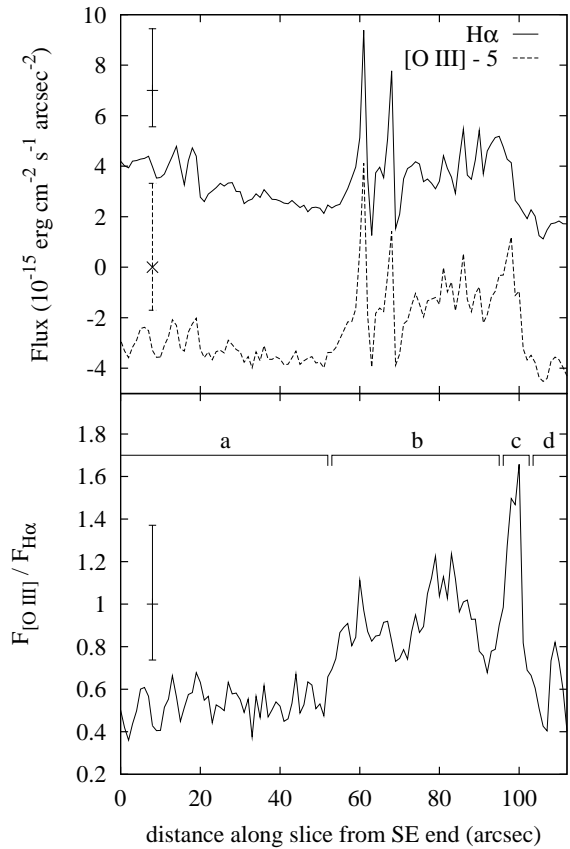}
\caption{The emission line flux (top panel) and the \ofig / H$\alpha$ ratio (lower panel) along a slice through the nebula indicated in Fig. 5. The \ofig flux has been offset by $-5\times 10^{-15}$ erg s$^{-1}$cm$^{-2}$arcsec$^{-2}$ for clarity. In the lower panel, region a = inside the nebula; b = the ring nebula; c = the thin outer shell luminous in \ofig / H$\alpha$; and d = outside the nebula. The mean error associated with each data set (equations 1--3) is indicated on the left in the panels. The spikes in the upper panel are due to stars in the slice.
}
\end{figure}

\subsection{Constraining the shock velocity and jet power}

We measured the $F_{\rm[O~\small III\normalsize ]}$ / $F_{\rm H\alpha}$ flux ratio in seven `slices' (typical thickness 12 pixels $= 4$ arcsec) of the [O~\small III\normalsize ]-emitting shock front, and in four larger regions behind it. In the thin outer shell we obtain $F_{\rm[O~\small III\normalsize ]}$ / $F_{\rm H\alpha}=1.36\pm 0.28$ and behind the shell, $F_{\rm[O~\small III\normalsize ]}$ / $F_{\rm H\alpha}=0.89\pm 0.46$. Taking into account the error from equation 3, $F_{\rm[O~\small III\normalsize ]}$ / $F_{\rm H\alpha}=1.36^{+0.65}_{-0.46}$ in the shock front and $F_{\rm[O~\small III\normalsize ]}$ / $F_{\rm H\alpha}=0.89^{+0.64}_{-0.49}$ in the nebula behind the shock front.

Both radiative and non-radiative shock waves are expected to produce a mix of line (radiative recombination) and continuum (due to bremsstrahlung emission) output \citep[e.g.][]{krol99}. Non-radiative shock waves (defined as when the cooling time of the shocked gas exceeds the age of the shock) produce Balmer optical emission lines only, because the heavier atoms are not excited \citep[e.g.][]{mckeho80}, so the existence of the strong [O~\small III\normalsize ] line confirms the radiative nature of the Cyg X--1 nebula shock. In a radiative bow shock, the ambient ISM gas is first perturbed by the approaching radiation field. The ionisation of preshock hydrogen becomes progressively more significant in the velocity range $90\leq v_{\rm s} \leq 120$ km s$^{-1}$; below 90 km s$^{-1}$ the preshock gas is essentially neutral and above 120 km s$^{-1}$ it is entirely ionised \citep[e.g.][]{shulmc79}. In addition, at velocities $\sim 100$ km s$^{-1}$ the ionising He~\small II\normalsize~(304\AA) photons from the cooling region can doubly ionise the oxygen increasing the intensity of the [O~\small III\normalsize ] lines \citep{mckeho80}. This leads to the high excitation line [O~\small III\normalsize ] becoming negligible at low velocities in all shock-wave models (\citealt*{raym79,shulmc79,dopiet84,bineet85,coxra85,hartet87,dopsu95}; 1996). The outer shell of the Cyg X--1 nebula, bright in [O~\small III\normalsize ] / H$\alpha$, may originate in the ionised atoms close to the front of the bow shock, and its presence confirms its association with the jet of Cyg X--1.

Radiative shock models predict that relative optical line strengths of the shocked gas are highly sensitive to the temperature and velocity of the shock wave, and vary only weakly with the conditions in the preshock gas such as the elemental abundances \citep[e.g.][]{coxra85}. Temperatures and velocities of shocked matter, mainly in SNRs, are commonly inferred by comparing their observed optical or ultraviolet (UV) line ratios to these models \citep[e.g.][]{blaiet91,vancet92,leveet95,boccet00,mavret01}. The bright [O~\small III\normalsize ] / H$\alpha$ shock front of the Cyg X--1 nebula appears to be thicker on the western side of the nebula, and the H$\alpha$ and [O~\small III\normalsize ] surface brightness is also higher in this region (Fig. 4). This implies the shock may be faster here, possibly due to a lower preshock density; it is the furthest area of the nebula from the H~\small II \normalsize region (or due to the proper motion of Cyg X--1; see below).

\begin{figure}
\centering
\includegraphics[height=8.6cm,angle=270]{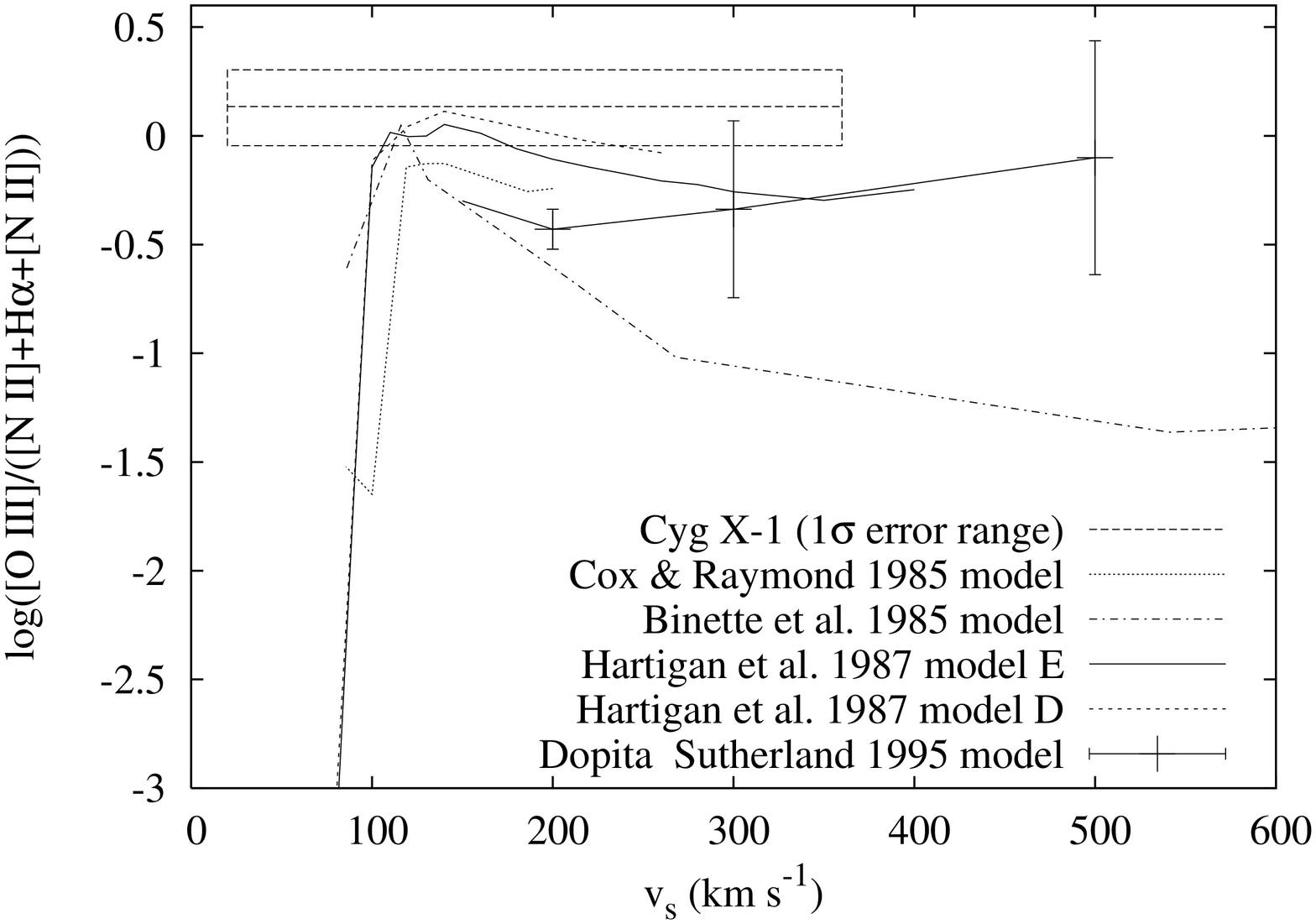}
\caption{The \ofig / H$\alpha$ ratio as a function of shock velocity predicted from five radiative shock models. The dashed rectangle represents the constrained range of values measured for the Cyg X--1 shock front \citep[velocity range from][]{gallet05}. The INT \ofig filter includes both lines of the doublet; 4958.9$\AA$  and 5006.9$\AA$, and the H$\alpha$ filter includes the two [N~II] lines at 6548.1$\AA$ and 6583.4$\AA$ in addition to 6562.8$\AA$ H$\alpha$, within its bandpass. We therefore plot log($F_{\rm \ofig+\ofig}$/$F_{\rm [N~II]+H\alpha +[N~II]}$) versus velocity.
}
\end{figure}

The relation between the shock velocity and [O~\small III\normalsize ] / H$\alpha$ ratio in the shell is described for most of the published models of steady-flow radiative shocks with self-consistent pre-ionisation \citep{coxra85,bineet85,hartet87,dopsu95}.  The reliability of these models can be assessed from their level of agreement, unless there are systematic uncertainties common to all models \citep[which seems not to be the case as the velocities predicted by the models are consistent with velocities observed using other methods; e.g.][]{michet00,punet02}. In Fig. 7 we plot the aforementioned relation predicted from five of the most advanced radiative shock models. We overplot the empirical [O~\small III\normalsize ] / H$\alpha$ ratio found for the Cyg X--1 shock front. Bearing in mind the previously estimated range of shock velocities, 20 $< v_{\rm s} < 360$ km s$^{-1}$, we find that outside the range $100\leq v_{\rm s} \leq 360$ km s$^{-1}$, no models successfully predict the observed [O~\small III\normalsize ] / H$\alpha$ measurement to within 1$\sigma$. Subsequently, we infer a temperature of the gas in the shocked shell \citep{gallet05} of $2.4\times 10^5 \leq T_{\rm s} \leq 3.1\times 10^6$ K. The models in Fig. 7 assume typical Galactic abundances and preshock particle densities within the range inferred for the Cyg X--1 nebula; $1\leq n_0 \leq 300$ cm$^{-3}$ \citep{gallet05}.

By constraining the velocity and temperature of the gas in the shock front, we are able to refine a number of parameters such as the particle density in the shell and power in the jet \citep{gallet05}. Table 2 lists the parameters inferred for the nebula and jet. We adopt the same methodology to calculate the parameters as used in \cite{gallet05}. The temperature-dependent ionisation fraction of the gas $x$ is taken from \cite{spit78}. At the temperatures inferred, the hydrogen in the shocked shell is fully ionised. We have improved the accuracy of the shock velocity, the time-averaged jet power and the jet lifetime from $> 1$ dex to $\sim 0.5$ dex. Assuming the power originates in the hard state jet which is switched on for $\sim 90$ percent of the lifetime of Cyg X--1 \citep[e.g.][]{gallet05}, the total power of both these jets is between 30 and 100 percent of the bolometric 0.1--200 keV X-ray luminosity \citep[$L_{\rm X}\approx 3\times 10^{37}$ erg s$^{-1}$;][]{disaet01}.

Assuming the shock velocity has been approximately constant over the lifetime of the jet, the jet lifetime $t\sim 0.02$--0.04 Myr is much shorter than the estimated age of the progenitor of the black hole, $\sim 5$--7 Myr \citep{miraro03}. However, given that the proper motion of Cyg X--1 is $\sim 8.7$ mas year$^{-1}$ in a direction $\sim 130$ degrees from the jet direction \citep{stiret01,lestet99,miraro03}, the system must have travelled between 2.5 and 9.1 arcmins during the lifetime of the jet. We speculate that the jet may have been present for the lifetime of Cyg X--1, but only when the system moved into the ISM close to the H~\small II \normalsize region did the local density become large enough for a bow shock to form. In fact, the age of the nebula (if it is powered by the jet) cannot exceed $\sim 0.04$ Myr because the position of Cyg X--1 at the birth of the jet would be \emph{in front of} the bow shock. This scenario may also explain the apparent higher [O~\small III\normalsize ] / H$\alpha$ ratio and velocity of the gas on the western side of the nebula compared with the eastern side: the western side is further from the jet origin as the Cyg X--1 system moves.

\begin{table}
\small
\caption{Revised nebula and jet parameters ($1\sigma$ confidence limits where quoted).}
\begin{tabular}{ll}
\hline
Parameter&Value\\
\hline
\multicolumn{2}{c}{Unchanged parameters \citep{gallet05}:}\vspace{2mm}\\
Monochromatic radio&$L_{\rm \nu, 1.4GHz}\approx 10^{18}$ erg s$^{-1}$Hz$^{-1}$\\
~~~~luminosity&\\
Cyg X--1 -- shell separation&$L\approx 3.3\times 10^{19}$ cm\\
Jet inclination angle&$\theta \approx 35^o$\\
~~~~to line of sight&\\
Thickness of shell&$\Delta R\approx 1.6\times 10^{18}$ cm\\
Source unit volume&$V(L,\Delta R)\approx 4\times 10^{53}$ cm$^3$\\
Bremsstrahlung emissivity&$\epsilon_{\nu}(V,L_{\rm \nu, 1.4GHz})\approx 2.5\times 10^{-36}$\\
~~~~for hydrogen gas&~~~~erg s$^{-1}$Hz$^{-1}$ cm$^{-3}$\\
Gaunt factor&$g\approx 6$\\
Number density of&$1\leq n_0 \leq 300$ cm$^{-3}$\\
~~~~preshock gas&\vspace{2mm}\\
\multicolumn{2}{c}{Refined parameters:}\vspace{2mm}\\
H$\alpha$ flux of shell&$F_{\rm H\alpha}\approx 3.6\times 10^{-15}$\\
&~~~~erg s$^{-1}$cm$^{-2}$arcsec$^{-2}$\\
H$\alpha$ luminosity of nebula&$1.8\times 10^{34}\leq L_{\rm H\alpha}\leq$\\
&~~~~$2.8\times 10^{35}$ erg s$^{-1}$\\
\ofig flux of shell&$F_{\rm[O~\small III\normalsize ]}\approx 2.8\times 10^{-15}$\\
&~~~~erg s$^{-1}$cm$^{-2}$arcsec$^{-2}$\\
\ofig luminosity of&$1.3\times 10^{34}\leq L_{\rm[O~\small III\normalsize ]}\leq$\\
~~~~nebula&~~~~$2.1\times 10^{35}$ erg s$^{-1}$\\
\ofig / H$\alpha$ ratio&$F_{\rm[O~\small III\normalsize ]}$ / $F_{\rm H\alpha}=1.36^{+0.65}_{-0.46}$\\
~~~~in shock front&\\
\ofig / H$\alpha$ ratio&$F_{\rm[O~\small III\normalsize ]}$ / $F_{\rm H\alpha}=0.89^{+0.64}_{-0.49}$\\
~~~~behind shock front&\\
Optical continuum flux&$F_{\rm \nu,OPT}\leq 6.8\mu$Jy arcsec$^{-2}$\\
~~~~density of shell&~~~~(3$\sigma$)\\
Shock velocity&$100 \leq v_{\rm s} \leq 360$ km s$^{-1}$\\
Shocked gas temperature&$2.4\times 10^5 \leq T_{\rm s} \leq 3.1\times 10^6$ K\\
Number density of ionised&$56\leq n_{\rm e}\leq 103$ cm$^{-3}$\\
~~~~particles in shell&\\
Ionisation fraction&$x\approx 1$\\
Number density of total&$56\leq n_{\rm p}\leq 103$ cm$^{-3}$\\
~~~~particles in the shell&\\
Jet lifetime&$0.017\leq t\leq 0.063$ Myr\\
Time-averaged jet power&$P_{\rm Jet} = (4$--$14) \times 10^{36}$ erg s$^{-1}$\\
Hard state jet power&$P_{\rm Jets} = (9$--$30) \times 10^{36}$ erg s$^{-1}$\\
~~~~including both jets&\\
Outflow power / X-ray&$0.3\leq f_{\rm Jet/X}\leq 1.0$\\
~~~~luminosity&\\
\hline
\end{tabular}
\normalsize
\end{table}

\subsection{Ruling out alternative origins of the shell}

We have shown (Section 4.1) that the morphology of the Cyg X--1 nebula indicates the gas is shock-excited and not photoionised; the structure is therefore not a planetary nebula or an H~\small II \normalsize region. The nebula also cannot be a SNR associated with Cyg X--1 itself \citep{gallet05}. The proximity of the nebula to Cyg X--1, its alignment with the Cyg X--1 jet and the agreement of its age inferred from its velocity and from the proper motion of Cyg X--1 (Section 4.2) all point towards the jet-powered scenario. The one remaining alternative is a field SNR not associated with Cyg X--1; we here explore this possibility.

\cite*{xuet05a} plot the diameter of 185 Galactic SNRs against their 1 GHz surface brightness (Fig. 3 of that paper). The radio surface brightness of the Cyg X--1 nebula is $\Sigma_{\rm 1~GHz}\approx 3\times 10^{-23}$ W m$^{-2}$ Hz$^{-1}$ sr$^{-1}$ \citep[lower than any SNR in the sample of][]{xuet05a}, and would have a diameter of $\ga 100$ pc if it were to lie close to the diameter--surface brightness relation for SNRs. The nebula has an angular diameter of $\sim 7$ arcmin, suggesting a distance to the source of $d \ga 50$ kpc if it is a SNR. However, the authors note that this surface brightness--diameter relation could be severely biased by selection effects \citep*{xuet05b}. Large, low surface-brightness SNRs may not have been detected in surveys due to confusion with strong foreground and background sources \citep[e.g.][]{gree04}. We therefore cannot completely rule out a SNR origin to the nebula, but its alignment with the relativistic jet of Cyg X--1 (for example) strongly suggests it is related to the system.

\begin{figure*}
\includegraphics[width=15cm,angle=0]{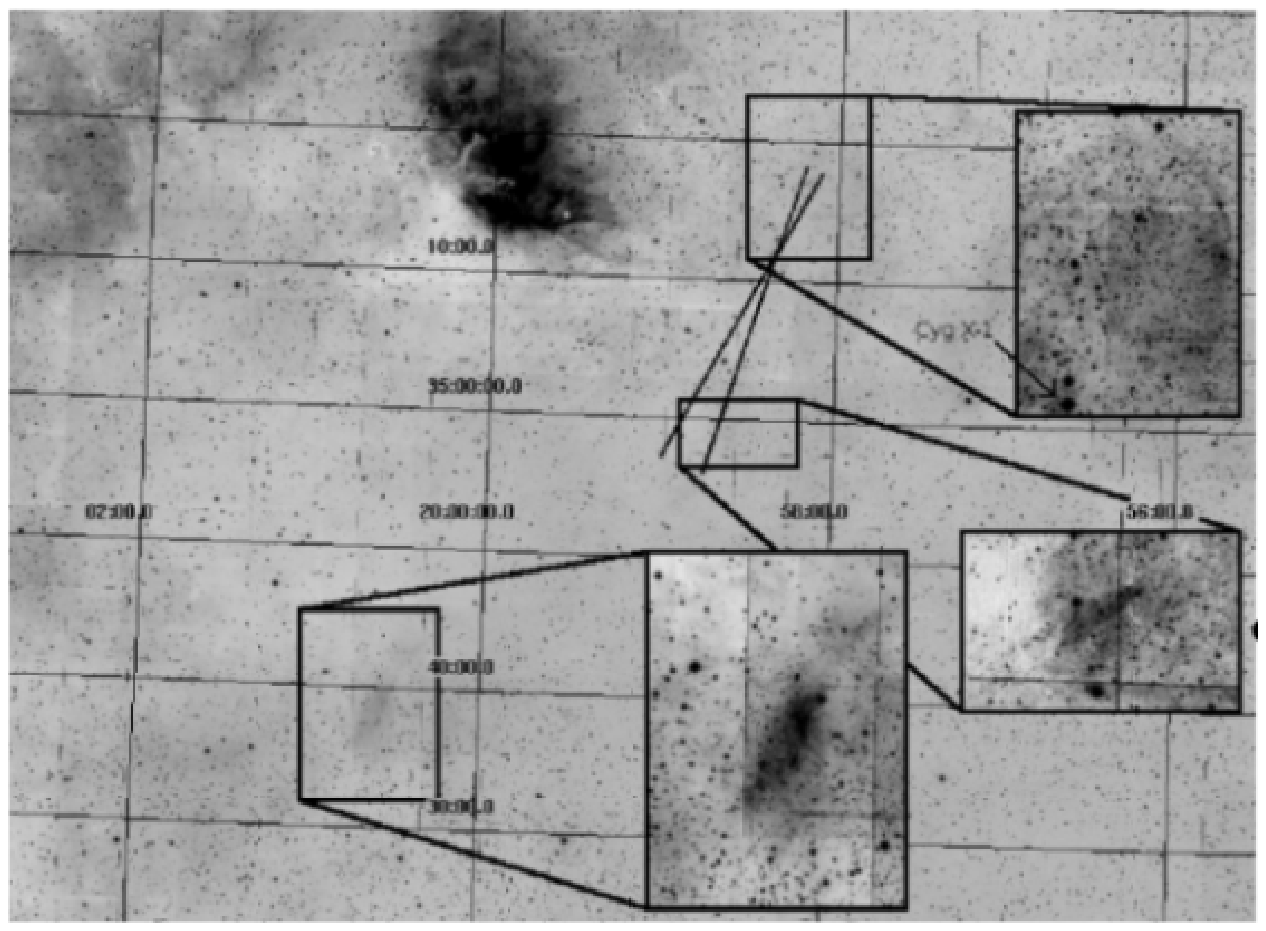}
\caption{A $\sim 1.5$ degree$^2$ mosaic from the IPHAS survey. Each CCD frame is a 120 sec H$\alpha$ exposure. North is to the top, east to the left. RA and Dec are epoch J2000. The most extreme opening position angles of the resolved radio jet (both the steady low/hard state and transient jets) north of Cyg X--1; $17^o\leq \theta \leq 28^o$ \citep[][also projected to the south assuming the jets are anti-parallel]{stiret01,fendet06} are shown. The north nebula and two areas of candidate ISM interactions with the southern jet of Cyg X--1 are expanded.
}
\end{figure*}

\subsection{The search for ISM interactions with the southern jet}

We searched a $\sim 0.25$ deg$^2$ field in [O~\small III\normalsize ] south of Cyg X--1 for a shell powered by the southern jet (lower panel of Fig. 1). No shell was observed in the 1800 sec image, nor any visual nebulosity detected in the [O~\small III\normalsize ] images that is consistent with the southern jet of Cyg X--1 interacting with the ISM. The mass density of ISM may be much less to the south; the jet should travel essentially ballistically and undetected until being decelerated by a denser region. \cite{hein02} found that the rarity of jet-powered lobes in microquasars compared to AGN is likely due to the lower density environment. In addition, it is interesting to note that the X-ray-bright shocked edges surrounding AGN radio lobes are sometimes one-sided \citep{krafet03}. We searched a $\sim1$ deg$^2$ H$\alpha$ mosaic from the IPHAS survey \citep[][see Fig. 8]{drewet05}. Two candidate `hot spots' analogous to those associated with AGN jets were identified, and are expanded in the figure.

Radio emission exists in the region close to the northern hot spot according to \cite{martet96}, but it is uncertain whether it originates in point sources or extended emission. We inspected the southern region of the 1.4 GHz radio image obtained by \cite{gallet05} to search for diffuse emission. In a 1600$\times$1600 arcsec region close to the more northern H$\alpha$ hot spot (the southern hot spot is not within the radio field of view), the mean r.m.s. flux density is $4\times 10^{-4}$ Jy beam$^{-1}$, yielding a 3$\sigma$ upper limit to the (putative) radio hot spot in that region of 1.2 mJy beam$^{-1}$. Deeper radio observations are needed to test whether the candidate H$\alpha$ hot spots have radio counterparts, as would be expected if they are jet--ISM interaction sites. In addition, observations in [S~\small II\normalsize ] would test whether the emitting gas is shock-excited or photoionised, and deeper [O~\small III\normalsize ] observations may constrain the velocity of any shocked gas.

\section{Conclusions}

We have imaged the shell-like nebula supposedly powered by the jet of Cyg X--1 in narrowband optical emission line filters. The nebula is bright in H$\alpha$ and [O~\small III\normalsize ] and the morphology and line ratios indicate that it is a radiative shock wave with a velocity $v_{\rm s}\geq 100$ km s$^{-1}$. We rule out alternative interpretations of the structure such as photoionised gas, but cannot completely rule out an unrelated SNR origin. However, the morphology of the nebula and its alignment with the jet of Cyg X--1 strongly suggest that it is indeed powered by this jet. We have probably isolated for the first time the thin, hot, compressed gas close to the bow shock front of a nebula powered by an X-ray binary jet. The time-averaged power of the jet must be $P_{\rm Jet} = (4$--$14) \times 10^{36}$ erg s$^{-1}$ to power a shock wave of the inferred size, luminosity, velocity and temperature. If the steady hard state jet of Cyg X--1 (that is active for $\sim90$ percent of the time) is providing this power \citep[rather than the transient jet;][]{fendet06}, the total energy of both jets in the hard state is 0.3--1.0 times the bolometric X-ray luminosity. This provides the strongest evidence so far that at an accretion rate of $\sim 2$ percent Eddington, Cyg X--1 is in a state of approximate equipartition between radiative and kinetic output. Lower accretion rates should therefore be jet-dominated (\citealt{fendet03}; \citealt*{kordet06}). The inferred age of the nebula is consistent with the time Cyg X--1 has been close to the currently nearby H~\small II \normalsize region.

No nebula associated with the southern jet of Cyg X--1 is observed, although we detect two putative H$\alpha$ `hot spots' that require optical emission line and radio follow-up to confirm their nature. Optical and UV \citep[e.g.][]{blaiet91} spectra of the northern shell will provide invaluable insights into the properties of the shocked gas, and hence the jet.

\vspace{5mm}
\emph{Acknowledgements}
We would like to thank Mischa Schirmer for extensive help with the usage of the data reduction pipeline package \small THELI\normalsize. We also thank Elmar K\"ording, Sebastian Heinz and Manfred Pakull for stimulating discussions. This paper makes use of data from the Isaac Newton Telescope, operated on the island of La Palma by the Isaac Newton Group in the Spanish Observatorio del Roque de los Muchachos of the Instituto de Astrofisica de Canarias. EG is supported by NASA through Chandra Postdoctoral Fellowship Award PF5-60037, issued by the Chandra X-Ray Observatory Center, which is operated by the Smithsonian Astrophysical Observatory for and on behalf of NASA under contract NAS8-39073.


\begin{thebibliography}{natbib}
\bibitem[\protect\citeauthoryear{Binette, Dopita \& Tuohy}{Binette et al.}{1985}]{bineet85}Binette L., Dopita M. A., Tuohy I. R., 1985, ApJ, 297, 476
\bibitem[\protect\citeauthoryear{Blair, Kirshner \& Chevalier}{Blair et al.}{1982}]{blaiet82}Blair W. P., Kirshner R. P., Chevalier R. A., 1982, ApJ, 254, 50
\bibitem[\protect\citeauthoryear{Blair et al.}{1991}]{blaiet91}Blair W. P., et al., 1991, ApJ, 379, L33
\bibitem[\protect\citeauthoryear{Blandford \& K\"onigl}{1979}]{blanko79}Blandford R. D., K\"onigl A., 1979, ApJ, 232, 34
\bibitem[\protect\citeauthoryear{Bocchino et al.}{2000}]{boccet00}Bocchino F., Maggio A., Sciortino S., Raymond J., 2000, A\&A, 359, 316
\bibitem[\protect\citeauthoryear{Burbidge}{1959}]{burb59}Burbidge G. R., 1959, ApJ, 129, 849
\bibitem[\protect\citeauthoryear{Carilli, Perley \& Harris}{Carilli et al.}{1994}]{cariet94}Carilli C. L., Perley R. A., Harris D. E., 1994, MNRAS, 270, 173
\bibitem[\protect\citeauthoryear{Castor, McCray \& Weaver}{Castor et al.}{1975}]{castet75}Castor J., McCray R., Weaver R., 1975, ApJ, 200, L107
\bibitem[\protect\citeauthoryear{Celotti \& Fabian}{1993}]{celofa93}Celotti A., Fabian A. C., 1993, MNRAS, 264, 228
\bibitem[\protect\citeauthoryear{Chaty et al.}{2001}]{chatet01}Chaty S., Rodr\'{i}guez L. F., Mirabel I. F., Geballe T. R., Fuchs Y., Claret A., Cesarsky C. J., Cesarsky D., 2001, A\&A, 366, 1035
\bibitem[\protect\citeauthoryear{Corbel \& Fender}{2002}]{corbfe02}Corbel, S., Fender, R. P. 2002, ApJ, 573, L35
\bibitem[\protect\citeauthoryear{Corbel et al.}{2002}]{corbet02}Corbel S., Fender R. P., Tzioumis A. K., Tomsick J. A., Orosz J. A., Miller J. M., Wijnands R., Kaaret P., 2002, Sci, 298, 196
\bibitem[\protect\citeauthoryear{Cox}{1972}]{cox72}Cox, D. P., 1972, ApJ, 178, 143
\bibitem[\protect\citeauthoryear{Cox \& Raymond}{1985}]{coxra85}Cox D. P., Raymond J. C., 1985, ApJ, 298, 651
\bibitem[\protect\citeauthoryear{Croston, Kraft \& Hardcastle}{Croston et al.}{2006}]{croset06}Croston J. H., Kraft R. P., Hardcastle M. J., 2006, astro-ph/0610889
\bibitem[\protect\citeauthoryear{Di Salvo et al.}{2001}]{disaet01}Di Salvo T., Done C., \.Zycki P. T., Burderi L., Robba N. R., 2001, ApJ, 547, 1024
\bibitem[\protect\citeauthoryear{Dopita et al.}{1984}]{dopiet84}Dopita M. A., Binette L., Dodorico S., Benvenuti P., 1984, ApJ, 276, 653
\bibitem[\protect\citeauthoryear{Dopita \& Sutherland}{1995}]{dopsu95}Dopita M., Sutherland R. S., 1995, ApJ, 455, 468
\bibitem[\protect\citeauthoryear{Dopita \& Sutherland}{1996}]{dopsu96}Dopita M., Sutherland R. S., 1996, ApJS, 102, 161
\bibitem[\protect\citeauthoryear{Drew et al.}{2005}]{drewet05}Drew J. E., et al., 2005, MNRAS, 362, 753
\bibitem[\protect\citeauthoryear{Dunn, Fabian \& Celotti}{Dunn et al.}{2006}]{dunnet06}Dunn R. J. H., Fabian A. C., Celotti A., 2006, MNRAS, 372, 1741
\bibitem[\protect\citeauthoryear{Erben et al.}{2005}]{erbeet05}Erben T., et al., 2005, AN, 326, 432
\bibitem[\protect\citeauthoryear{Fender}{2001}]{fend01}Fender R. P., 2001, MNRAS, 322, 31
\bibitem[\protect\citeauthoryear{Fender}{2006}]{fend06}Fender R. P., 2006, in Compact Stellar X-Ray Sources, eds. Lewin W. H. G., van der Klis M., Cambridge University Press, p. 381
\bibitem[\protect\citeauthoryear{Fender, Belloni \& Gallo}{Fender et al.}{2004}]{fendet04}Fender, R. P., Belloni, T. M., Gallo, E. 2004, MNRAS, 355, 1105
\bibitem[\protect\citeauthoryear{Fender, Gallo \& Jonker}{Fender et al.}{2003}]{fendet03}Fender R. P., Gallo E., Jonker P. G., 2003, MNRAS, 343, L99
\bibitem[\protect\citeauthoryear{Fender et al.}{2006}]{fendet06}Fender R. P., Stirling A. M., Spencer R. E., Brown I., Pooley G. G., Muxlow T. W. B., Miller-Jones J. C. A., 2006, MNRAS, 369, 603
\bibitem[\protect\citeauthoryear{Fomalont, Geldzahler \& Bradshaw}{Fomalont et al.}{2001}]{fomaet01}Fomalont E. B., Geldzahler B. J., Bradshaw C. F., 2001, ApJ, 558, 283
\bibitem[\protect\citeauthoryear{Forman et al.}{2005}]{formet05}Forman W., et al., 2005, ApJ, 635, 894
\bibitem[\protect\citeauthoryear{Gallo, Fender \& Pooley}{Gallo et al.}{2003}]{gallet03}Gallo, E., Fender, R. P., Pooley, G. G. 2003, MNRAS, 344, 60
\bibitem[\protect\citeauthoryear{Gallo et al.}{2005}]{gallet05}Gallo E., Fender R. P., Kaiser C., Russell, D. M., Morganti R., Oosterloo T., Heinz S., 2005, Nat, 436, 819
\bibitem[\protect\citeauthoryear{Green}{2004}]{gree04}Green D. A., 2004, Bull. Astr. Soc. India, 32, 335
\bibitem[\protect\citeauthoryear{Hartigan, Raymond \& Hartmann}{Hartigan et al.}{1987}]{hartet87}Hartigan P., Raymond J., Hartmann L., 1987, ApJ, 316, 323
\bibitem[\protect\citeauthoryear{Hartmann \& Burton}{1997}]{hartbu97}Hartmann D., Burton W. B., 1997, Atlas of Galactic Neutral Hydrogen, Cambridge Univ. Press, Cambridge
\bibitem[\protect\citeauthoryear{Heinz}{2002}]{hein02}Heinz S., 2002, A\&A, 388, L40
\bibitem[\protect\citeauthoryear{Heinz}{2006}]{hein06}Heinz S., 2006, ApJ, 636, 316
\bibitem[\protect\citeauthoryear{Heinz, Reynolds \& Begelman}{Heinz et al.}{1998}]{heinet98}Heinz S., Reynolds C. S., Begelman M. C., 1998, ApJ, 501, 126
\bibitem[\protect\citeauthoryear{Homan \& Belloni}{2005}]{homabe05}Homan J., Belloni T., 2005, Ap\&SS, 300, 107
\bibitem[\protect\citeauthoryear{Homan et al.}{2005}]{homaet05}Homan, J., Buxton, M., Markoff, S., Bailyn, C. D., Nespoli, E., Belloni, T. 2005, ApJ, 624, 295
\bibitem[\protect\citeauthoryear{Johnson}{1953}]{john53}Johnson H. M., 1953, ApJ, 118, 370
\bibitem[\protect\citeauthoryear{Kaiser \& Alexander}{1997}]{kaisal97}Kaiser C. R., Alexander P, 1997, MNRAS, 286, 215
\bibitem[\protect\citeauthoryear{Kaiser et al.}{2004}]{kaiset04}Kaiser C. R., Gunn K. F., Brocksopp C., Sokoloski J. L., 2004, ApJ, 612, 332
\bibitem[\protect\citeauthoryear{King}{1985}]{king85}King D. L., 1985, ING La Palma Technical Notes, 31
\bibitem[\protect\citeauthoryear{K\"ording, Fender \& Migliari}{K\"ording et al.}{2006}]{kordet06}K\"ording E., Fender R. P., Migliari S., 2006, MNRAS, 369, 1451
\bibitem[\protect\citeauthoryear{Kraft et al.}{2003}]{krafet03}Kraft R. P., V\'azquez S. E., Forman W. R., Jones C., Murray S. S., Hardcastle M. J., Worrall D. M., Churazov E., 2003, ApJ, 592, 129
\bibitem[\protect\citeauthoryear{Krolik}{1999}]{krol99}Krolik J. H., 1999, Active Galactic Nuclei: From the Central Black Hole to the Galactic Environment, Princeton Univ. Press, Princeton
\bibitem[\protect\citeauthoryear{Lestrade et al.}{1999}]{lestet99}Lestrade J.-F., Preston R. A., Jones D. L., Phillips R. B., Rogers A. E. E., Titus M. A., Rioja M. J., Gabuzda D. C., 1999, A\&A, 344, 1014
\bibitem[\protect\citeauthoryear{Levenson et al.}{1995}]{leveet95}Levenson N. A., Kirshner R. P., Blair W. P., Winkler P. F., 1995, AJ, 110, 739
\bibitem[\protect\citeauthoryear{Mart\'i et al.}{1996}]{martet96}Mart\'i J., Rodriguez L. F., Mirabel I. F., Paredes J. M., 1996, A\&A, 306, 449
\bibitem[\protect\citeauthoryear{Massey, Johnson \& Degioia-Eastwood}{Massey et al.}{1995}]{masset95}Massey P., Johnson K. E., Degioia-Eastwood K., 1995, ApJ, 454, 151
\bibitem[\protect\citeauthoryear{Mavromatakis et al.}{2001}]{mavret01}Mavromatakis F., Ventura J., Paleologou E. V., Papamastorakis J., 2001, A\&A, 371, 300
\bibitem[\protect\citeauthoryear{McClintock \& Remillard}{2006}]{mcclet06}McClintock J. E., Remillard R. A., 2006, in Compact Stellar X-Ray Sources, eds. Lewin W. H. G., van der Klis M., Cambridge University Press, p. 157
\bibitem[\protect\citeauthoryear{McKee \& Hollenbach}{1980}]{mckeho80}McKee C. F., Hollenbach D. J., 1980, ARA\&A, 18, 219
\bibitem[\protect\citeauthoryear{Michael et al.}{2000}]{michet00}Michael E., et al., 2000, ApJ, 542, L53
\bibitem[\protect\citeauthoryear{Migliari, Fender \& M\'endez}{Migliari et al.}{2002}]{miglet02}Migliari S., Fender R., M\'endez, M., 2002, Sci, 297, 1673
\bibitem[\protect\citeauthoryear{Mirabel et al.}{1992}]{miraet92}Mirabel I. F., Rodr\'{i}guez L. F., Cordier B., Paul J., Lebrun F., 1992, Nat, 358, 215
\bibitem[\protect\citeauthoryear{Mirabel \& Rodrigues}{2003}]{miraro03}Mirabel I. F., Rodrigues I., 2003, Sci, 300, 1119
\bibitem[\protect\citeauthoryear{Nowak et al.}{2005}]{nowaet05}Nowak M. A., Wilms J, Heinz S., Pooley G., Pottschmidt K., Corbel S., 2005, ApJ, 626, 1006
\bibitem[\protect\citeauthoryear{Nulsen et al.}{2005}]{nulset05}Nulsen P. E. J., McNamara B. R., Wise M. W., David L. P., 2005, ApJ, 628, 629
\bibitem[\protect\citeauthoryear{Ogley et al.}{2000}]{ogleet00}Ogley R. N., Bell Burnell S. J., Fender R. P., Pooley G. G., Waltman E. B., 2000, MNRAS, 317, 158
\bibitem[\protect\citeauthoryear{Osterbrock}{1989}]{oste89}Osterbrock D. E., 1989, Astrophysics of Gaseous Nebulae and Active Galactic Nuclei. University Science Books, Sausalito CA
\bibitem[\protect\citeauthoryear{Perryman et al.}{1997}]{perret97}Perryman M. A. C., et al., 1997, A\&A, 323, L49
\bibitem[\protect\citeauthoryear{Predehl \& Schmitt}{1995}]{predet95}Predehl, P., Schmitt, J. H. M. M. 1995, A\&A, 293, 889
\bibitem[\protect\citeauthoryear{Pun et al.}{2002}]{punet02}Pun C. S. J., et al., 2002, ApJ, 572, 906
\bibitem[\protect\citeauthoryear{Raymond}{1979}]{raym79}Raymond J. C., 1979, ApJS, 39, 1
\bibitem[\protect\citeauthoryear{Rodr\'{i}guez et al.}{1992}]{rodret92}Rodr\'{i}guez L. F., Mirabel I. F., Mart\'{i} J., 1992, ApJ, 401, L15
\bibitem[\protect\citeauthoryear{Russell et al.}{2006}]{russet06}Russell D. M., Fender R. P., Hynes R. I., Brocksopp C., Homan J., Jonker P. G., Buxton M. M., 2006, MNRAS, 371, 1334
\bibitem[\protect\citeauthoryear{Schwartz}{1983}]{schw83}Schwartz R. D., 1983, ARA\&A, 21, 209
\bibitem[\protect\citeauthoryear{Shull \& McKee}{1979}]{shulmc79}Shull J. M., McKee C. F., 1979, ApJ, 227, 131
\bibitem[\protect\citeauthoryear{Spitzer}{1978}]{spit78}Spitzer L., 1978, Physical processes in the interstellar medium, New York Wiley-Interscience, 333
\bibitem[\protect\citeauthoryear{Stirling et al.}{2001}]{stiret01}Stirling A. M., Spencer R. E., de la Force C. J., Garrett M. A., Fender R. P., Ogley R. N., 2001, MNRAS, 327, 1273
\bibitem[\protect\citeauthoryear{Tudose et al.}{2006}]{tudoet06}Tudose V., Fender R. P., Kaiser C. R., Tzioumis A. K., van der Klis M., Spencer R., 2006, MNRAS, 372, 417
\bibitem[\protect\citeauthoryear{Vancura et al.}{1992}]{vancet92}Vancura O., Blair W. P., Long K. S., Raymond J. C., 1992, ApJ, 394, 158
\bibitem[\protect\citeauthoryear{Wilson, Smith \& Young}{Wilson et al.}{2006}]{wilset06}Wilson A. S., Smith D. A., Young A. J., 2006, ApJ, 644, L9
\bibitem[\protect\citeauthoryear{Wu et al.}{1982}]{wuet82}Wu C.-C., Holm A. V., Eaton J. A., Milgrom M., Hammerschlag-Hensberge G., 1982, PASP, 94, 149
\bibitem[\protect\citeauthoryear{Xu, Zhang \& Han}{Xu et al.}{2005a}]{xuet05a}Xu J-W., Zhang X-Z., Han J-L., 2005a, ChJAA, 5, 165
\bibitem[\protect\citeauthoryear{Xu, Zhang \& Han}{Xu et al.}{2005b}]{xuet05b}Xu J-W., Zhang X-Z., Han J-L., 2005b, ChJAA, 5, 442

\end{thebibliography}
\end{document}